\documentclass[
    ,final            
  ]
  {aipproc}

\layoutstyle{6x9}

\newcommand{\beq}{\begin{eqnarray}}
\newcommand{\eeq}{\end{eqnarray}}

\begin{document}

\title{Single Transverse-Spin Asymmetry in\\ Open Charm Production in SIDIS}


\author{Hiroo Beppu}{
  address={Graduate School of Science and Technology, Niigata University,
Ikarashi, Niigata 950-2181, Japan}
}

\author{Yuji Koike}{
  address={Department of Physics, Niigata University, Ikarashi, Niigata 950-2181, Japan}
}

\author{Kazuhiro Tanaka}{
  address={Department of Physics, Juntendo University, Inzai, Chiba 270-1695, Japan}
}

\author{Shinsuke Yoshida}{
  address={Graduate School of Science and Technology, Niigata University,
Ikarashi, Niigata 950-2181, Japan}
}
\begin{abstract}
We discuss the single transverse-spin asymmetry (SSA)
to be observed in the $D$-meson production with large transverse-momentum
in semi-inclusive DIS (SIDIS), $e p^\uparrow  \rightarrow  e D X$.
This contribution is embodied as a twist-3 mechanism in the collinear factorization,
which is induced by purely gluonic correlation inside the transversely-polarized nucleon,
in particular, by the three-gluon correlation effects.
We derive the first complete formula for the corresponding SSA in the leading-order QCD,
revealing the five independent structures with respect to the dependence
on the azimuthal angle for the produced $D$-meson.
Our result obeys universal structure behind the SSAs
in a variety of hard processes.
We present the numerical calculations of our SSA formula for the $D$-meson production
at the kinematics relevant to a future Electron Ion Collider.

\end{abstract}

\maketitle


Charm productions in semi-inclusive DIS (SIDIS), as well as in the $pp$ collision, are known to
be associated with the twist-2 gluon distributions in the nucleon, since
the $c\bar{c}$-pair creation through the photon-gluon or gluon-gluon fusion
is their driving subprocess. 
Similarly, the twist-3 contributions in the charm productions can be generated by
the purely gluonic effects inside the nucleon, in particular, the multi-gluon correlations.
Indeed, the observation of 
the single transverse-spin asymmetry (SSA) in the open charm productions 
allows us to probe the corresponding 
twist-3 effects~\cite{Kang:2008qh,Kang:2008ih,Beppu:2010qn,Koike:2011ns,Koike:2010jz}.

The corresponding SSA
arises as a {\em naively T-odd} effect 
in the cross section for the scattering of transversely-polarized nucleon 
off an unpolarized particle,
observing a 
$D$-meson 
with momentum $P_h$ 
in the final state, and this requires,
(i) nonzero transverse-momentum $P_{h\perp}$ originating 
from transverse motion
of 
quark or gluon; 
(ii) nucleon helicity flip in the cut diagrams for the cross section, corresponding
to the transverse polarization; 
and (iii) interaction beyond Born level to produce the
interfering phase between the LHS and the RHS of the cut in those diagrams. 
In particular, 
for large $P_{h\perp}\gg \Lambda_{\rm QCD}$ to be dealt with 
the collinear-factorization framework,
(i) should come from perturbative mechanism
as the recoil from the hard (unobserved) final-state partons, 
while nonperturbative effects 
can participate in the other two, (ii) and (iii), allowing us to obtain observably large SSA.
This 
is realized 
with the twist-3 multi-gluon correlation functions 
for the transversely-polarized nucleon~\cite{Beppu:2010qn}.
This twist-3 mechanism may be considered as an extension of the corresponding mechanism 
for
the SSA in the pion productions in the SIDIS~\cite{EKT07}, $pp$ collisions~\cite{To}, etc., 
based on the quark-gluon correlations in the nucleon,
but it has been clarified~\cite{Beppu:2010qn} that a straightforward extension~\cite{Kang:2008qh,Kang:2008ih} 
leads to missing many terms in the SSA.
We shall discuss the leading-order (LO) QCD result for the SSA 
in the high-$P_{h\perp}$ $D$-meson production
in SIDIS, $e(\ell)+p(p,S_\perp)\to e(\ell')+D(P_h)+X$, controlled by the twist-3 
three-gluon correlation functions in the nucleon. We also present a numerical estimate of the corresponding 
SSA with kinematics of a future Electron Ion Collider (EIC).
We use, as usual, the kinematic variables $S_{ep}=(\ell+p)^2$, $q=\ell-\ell'$, $Q^2=-q^2$, $x_{bj}=Q^2/(2p\cdot q)$, 
and $z_f=p\cdot P_h/ (p\cdot q)$.
We work in a frame where the momenta $\vec{q}$ and $\vec{p}$ are collinear, 
and define $q_T \equiv P_{h\perp}/z_f$ and the azimuthal angles $\phi$, $\Phi_S$, and $\chi$
of the lepton plane, the spin vector $S_\perp^\mu$, and the $D$-meson momentum $P_h^\mu$, 
respectively~\cite{Beppu:2010qn,Koike:2011ns}.
We take into account the masses $m_c$ and $m_h$ for the charm quark and the $D$ meson.

\begin{figure}
\includegraphics[height=3.6cm]{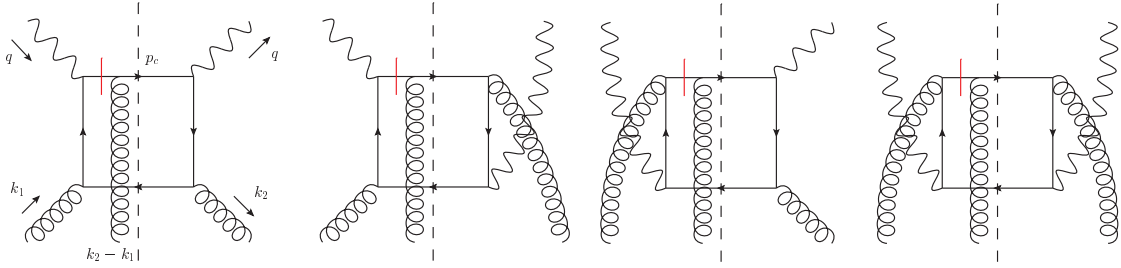}
\caption{Feynman diagrams for partonic subprocess in $ep^\uparrow\to eDX$;
mirror diagrams also contribute.}
\label{fig:1}
\end{figure}
In the LO in QCD perturbation theory, 
the photon-gluon fusion subprocesses of Fig.~1 
drive the SSA,
where the above-mentioned (i) is provided by the recoil from the hard unobserved $\bar c$ quark
and the $c$ quark with the momentum $p_c$ fragments into the $D$-meson in the final state.
The short bar on the internal $c$-quark line
indicates that the pole part is to be taken from that propagator,
to produce the
interfering phase of~(iii);
we note that these pole contributions from Fig.~1 would cancel the similar contributions 
from the corresponding mirror diagrams, if the $c$ quark were unobserved in
the final state as in the $\bar D$-meson production.
The external curly lines represent the gluons that are generated from the three-gluon
correlations present inside the transversely-polarized nucleon, 
$\langle p S_\perp| A_{\alpha}(0)A_{\beta}(\eta) A_{\gamma}(\xi)|p S_\perp \rangle$, corresponding to (ii). 
The diagrams obtained by the permutation of the gluon lines in Fig.~1 also satisfy (i)-(iii), but
the Bose statistics of the gluons in the above matrix element 
guarantees that we need not consider those diagrams separately.
Thus, the SSA in the present context 
can be derived entirely as the contributions of soft-gluon-pole (SGP) type~\cite{EKT07}, leading to $k_2-k_1=0$, 
by evaluating the pole part in Fig.~1. 
The twist-3 nature of those contributions are unraveled
by the collinear expansion, as usual.
The expansion produces lots of terms, 
each of which is not gauge invariant.
Indeed, many of them vanish or cancel eventually, and the remaining terms can be 
organized into a gauge-invariant form.
This can be demonstrated~\cite{Beppu:2010qn} by 
sophisticated use of the Ward identities
for the contributions of the diagrams in Fig.~1.
The resulting factorization formula of the spin-dependent, differential cross section
for $ep^\uparrow\to eDX$ reads~\cite{Beppu:2010qn,Koike:2011ns}
($[d\omega] \equiv dx_{bj}dQ^2dz_fdq_T^2d\phi d\chi$, 
for the differential elements)
\beq
&&
\hspace{-0.7cm}
\frac{d^6\Delta\sigma
}{[d\omega]}
=\frac{\alpha_{em}^2\alpha_se_c^2 M_N}{16\pi^2
 x_{bj}^2S_{ep}^2Q^2}\! \left(\frac{-\pi}{2}\right) \!
\sum_{k}
{\cal
 A}_k{\cal S}_k 
\int
\frac{dx}{x}\!
\int\frac{dz}{z} 
\delta \! \left(
\frac{q_T^2}{Q^2}-\! \left(1-\frac{1}{\hat{x}}\right) \! \left(1-\frac{1}{\hat{z}}\right)\!
+\frac{m_c^2}{\hat{z}^2Q^2}\right)
\nonumber\\
&&
\!\!\!\!\!\!\!\!
\times\! 
\sum_{a=c,\bar{c}}
D_a(z) 
\left(\!\delta_a \!\left\{
\left[\frac{dO(x,x)}{dx}-\frac{2O(x,x)}{x}\right]\! \Delta\hat{\sigma}^{1}_k
+\!\left[\frac{dO(x,0)}{dx}-\frac{2O(x,0)}{x}\right]\! \Delta\hat{\sigma}_k^2
\right. \right.
\nonumber\\
&&
\!\!\!\!\!\!\!\!\!\!\!\!
\left.\left.\left. \left.
+\!\frac{O(x,x)}{x}\Delta\hat{\sigma}^{3}_k
+\!\frac{O(x,0)}{x}\Delta\hat{\sigma}^{4}_k
\right\} 
+\right\{ O(x,x)\rightarrow N(x,x),\;  O(x,0)\rightarrow - N(x,0) \right\} \right)\! ,
\label{3gluonresult}
\eeq
where $\hat{x}=x_{bj}/x$, $\hat{z}=z_f/z$, and
a complete set of twist-3 
gluonic correlation functions 
is defined through
the gauge-invariant lightcone correlation of three field-strength tensors~\cite{Beppu:2010qn}
($n^2=0$, $p\cdot n=1$),
\begin{eqnarray}
&&
\hspace{-0.7cm}
{\cal M}^{\alpha\beta\gamma}_{F}(x_1,x_2)\! \equiv \!
-gi^3\!\!\int{d\lambda\over 2\pi}\int{d\zeta\over 2\pi}e^{i\lambda x_1}
e^{i\zeta(x_2-x_1)}\langle p S_\perp|d^{bca}F_b^{\beta n}(0)F_c^{\gamma n}(\zeta n)F_a^{\alpha n}(\lambda n)
|p S_\perp\rangle
\nonumber\\
&&
\!\!\!\!\!
\!\!\!\!\!\!\!\!\!\!
=\! 2iM_N \!\!
\left[
O(x_1,x_2)g^{\alpha\beta}\epsilon^{\gamma pnS_\perp}
\!+\! O(-x_2,x_1-x_2)g^{\beta\gamma}\epsilon^{\alpha pnS_\perp}
\!+\! O( x_2-x_1, -x_1)g^{\gamma\alpha}\epsilon^{\beta pnS_\perp}\right]\! ,
\label{3gluonO}
\end{eqnarray}
and similarly for $N(x_1,x_2)$ with the replacement $d^{bca}\rightarrow if^{bca}$,
and $D_c(z)$ denotes the 
usual twist-2 fragmentation function for a $c$-quark to become the $D$-meson.
The quark-flavor index $a$ can, in principle, be $c$ and $\bar{c}$, with $\delta_c=1$ and  
$\delta_{\bar{c}}=-1$, so that the cross section for 
the $\bar{D}$-meson production $ep^\uparrow\to e\bar{D}X$
can be obtained by 
a simple replacement of the fragmentation function 
to that for the $\bar{D}$ meson, $D_a (z) \rightarrow \bar{D}_a (z)$. 
The summation $\sum_k$ implies that the subscript $k$ runs over $1,2,3,4,8,9$, with
${\cal A}_1=1+\cosh^2\psi$,
${\cal A}_2=-2$,
${\cal A}_3=-\cos\tilde{\phi}\sinh 2\psi$, 
${\cal A}_4=\cos 2\tilde{\phi}\sinh^2\psi$,
${\cal A}_8=-\sin\tilde{\phi}\sinh 2\psi$, 
and ${\cal A}_9=\sin 2\tilde{\phi}\sinh^2\psi$,
where $\cosh\psi \equiv 2x_{bj}S_{ep}/ Q^2 -1$ and $\tilde{\phi}=\phi-\chi$, 
and ${\cal S}_k$ is defined as
${\cal S}_k=\sin(\Phi_S-\chi)$ for $k=1,2,3,4$
and ${\cal S}_k=\cos(\Phi_S-\chi)$ for $k=8,9$.
$\alpha_{em}$
is the fine-structure constant,
and
$e_c=2/3$ is the electric charge of the $c$-quark.  
Partonic hard parts $\Delta\hat{\sigma}_k^i$ 
depend on $m_c$ as well as other partonic variables;
for the explicit formulae of $\Delta\hat{\sigma}_k^i$, we refer the readers to \cite{Beppu:2010qn}.
Note that, instead of evaluating the diagrams in Fig.~1 
as above,
those results can be obtained
using the ``master formula''~\cite{Koike:2011ns}, 
schematically given by
\begin{equation}
\frac{d^6\Delta\sigma
}{[d\omega]}
\sim -i\pi\int
\frac{dx}{x^2}\!
\int\frac{dz}{z} 
{\partial {\cal H}_{\mu\nu}(xp,q, p_c)
\over \partial p_{c\perp}^\sigma}\otimes
{\cal M}^{\mu\nu\sigma}_{F}(x,x) \otimes D_a(z),
\label{master}
\end{equation}
where
${\cal H}_{\mu\nu}(xp,q, p_c)$ denotes the
partonic hard part 
for the $2\to2$ Born subprocess, expressed by the diagrams in Fig.~1 with the
soft ($k_2-k_1=0$) gluon line removed.
This reveals that $\Delta\hat{\sigma}_k^i$ in (\ref{3gluonresult}) are related 
to the twist-2 hard parts ${\cal H}_{\mu\nu}(xp,q, p_c)$,
similarly as 
in the SSA in various processes 
associated with 
twist-3 quark-gluon correlation functions.

Reexpressing as $\phi-\chi =\tilde{\phi}= \phi_h$, $\Phi_S-\chi=\phi_h-\phi_S$ in
the above-mentioned explicit forms of ${\cal A}_k$ and ${\cal S}_k$,
where $\phi_h$ and $\phi_S$ represent the azimuthal angles of the hadron plane 
and 
the nucleon's spin vector $\vec{S}_{\perp}$, respectively, measured from the {\it lepton plane},
one may express (\ref{3gluonresult}) as the superposition of five sine 
modulations,
\begin{equation}
\frac{d^6\Delta \sigma}{[d\omega]}
=f_1\sin(\phi_h-\phi_S)
+f_2\sin(2\phi_h-\phi_S)
+f_3\sin\phi_S
+f_4\sin(3\phi_h-\phi_S)
+f_5\sin(\phi_h+\phi_S) ,
\label{azimuth2}
\end{equation}
with the corresponding structure functions $f_1,f_2,\ldots, f_5$, similarly as
the twist-3 SSA
for $ep^\uparrow\to e\pi X$, generated from the quark-gluon correlation functions~\cite{KT071}.

As an illustration, we evaluate the SSA for the $D^0$ production, $ep^\uparrow\to eD^0 X$, 
using (\ref{3gluonresult}), (\ref{azimuth2}).
In particular, we present $F_1 \equiv f_1/\sigma_1^{U}$ with 
the kinematics relevant to a future EIC in Fig.~2,
where $\sigma_1^{U}$ denotes the twist-2 unpolarized cross section for $ep \to eD^0X$,
$d^6 \sigma^{\rm unpol}/[d\omega]$, averaged and integrated over the azimuthal angles $\phi_h$ and $\chi$, respectively.
The solid and dashed curves show the contributions generated by the three-gluon correlation functions
of (\ref{3gluonO}), assuming
$O(x,x)=O(x,0)=0.005xG(x)$~\cite{Koike:2010jz} (see also \cite{Kang:2008qh,Kang:2008ih})
with the unpolarized gluon-density distribution $G(x)$ for the nucleon
and using CTEQ6L parton distributions and KKKS fragmentation functions~\cite{Kneesch:2007ey} 
with the scale $\mu^2 = Q^2+m_c^2 +z_f^2q_T^2$.
The results demonstrate good chance to access multi-gluon effects at EIC.
The detailed numerical studies, including those for the asymmetries generated by the other structures
$f_2, \ldots, f_5$ in (\ref{azimuth2}), will be presented elsewhere.

\begin{figure}
\includegraphics[height=7.4cm]{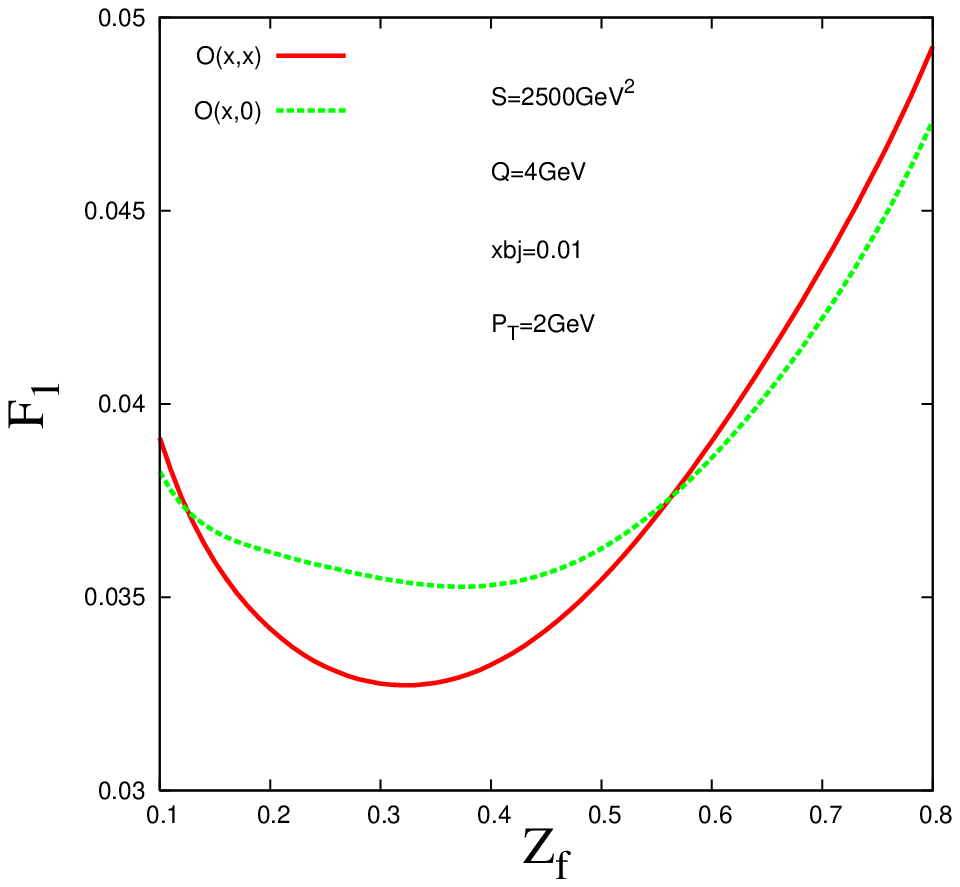}
\hspace{-1.5cm}
\includegraphics[height=7.4cm]{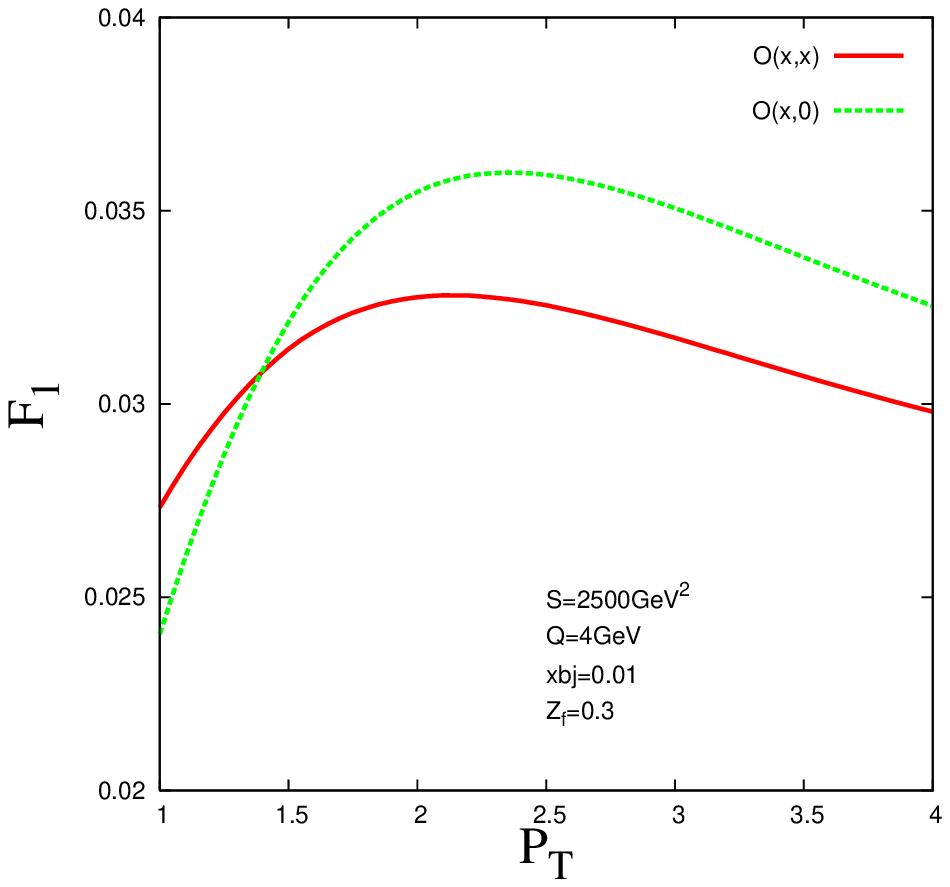}
\caption{The SSA as a function of (a)~$z_f$ and (b)~$P_T\equiv P_{h\perp}$ 
at EIC kinematics
with $S_{ep}=2500$~GeV$^2$.}
\label{fig:2}
\end{figure}


\medskip

The work of Y.K. and K.T. is supported by the Grant-in-Aid for Scientific Research 
No.23540292. 
The work of S.Y. is supported by the Grand-in-Aid for Scientific Research
(No.~22.6032) from the Japan Society of Promotion of Science.





\IfFileExists{\jobname.bbl}{}
 {\typeout{}
  \typeout{******************************************}
  \typeout{** Please run "bibtex \jobname" to optain}
  \typeout{** the bibliography and then re-run LaTeX}
  \typeout{** twice to fix the references!}
  \typeout{******************************************}
  \typeout{}
 }

\end{document}